\documentclass[aps, pre, twocolumn, amsmath, superscriptaddress,showkeys,showpacs]{revtex4-1}

\usepackage{commath}
\usepackage[innercaption]{sidecap}
\usepackage{amssymb}
\usepackage{graphicx}
\usepackage{graphicx,epstopdf}
\usepackage{xcolor, soul}
\usepackage{dcolumn}
\usepackage{bm}
\usepackage{mathrsfs} 
\usepackage{amsmath} 
\usepackage[colorlinks=true,linkcolor=blue,citecolor=blue]{hyperref}
\usepackage{fontenc}
\usepackage{float}
\usepackage{amsthm}
\usepackage{subfigure}
\usepackage{color}
\usepackage{ragged2e}
\usepackage{enumerate}
\usepackage{hyperref}
\usepackage[normalem]{ulem}

\def\be{\begin{equation}}
	\def\ee{\end{equation}}
\def\bea{\begin{eqnarray}}
	\def\eea{\end{eqnarray}}
\def\bfg{\begin{figure}[H]}
	\def\efg{\end{figure}}

\begin{document}

    \title{{Effect of mobility  in the  rock-paper-scissor dynamics with high mortality}}
	\author{Sahil Islam} 
	
	\email{thesahil.islam@gmail.com}
	
	\affiliation{Department of Physics, Jadavpur University, Jadavpur, Kolkata-700032}

	\author{Argha Mondal}
		\email{arghamondalb1@gmail.com}
	\affiliation{Department of Mathematics, Sidho-Kanho-Birsha University, Purulia-723104, WB, India}
	\affiliation{Department of Mathematical Sciences, University of Essex, Wivenhoe Park, UK}
	
	\author{Mauro Mobilia}
		\email{M.Mobilia@leeds.ac.uk}
	\affiliation{
		Department of Applied Mathematics, School of Mathematics, University of Leeds,
		Leeds LS2 9JT, United Kingdom}

	\author{Sirshendu Bhattacharyya} 
	\thanks{Corresponding Author}
	\email{sirs.bh@gmail.com}
	
	\affiliation{Department of Physics, Raja Rammohun Roy Mahavidyalaya,
		Radhanagar, Hooghly 712406, India}

	\author{Chittaranjan Hens}
	\thanks{Corresponding Author}
	\email{chittaranjanhens@gmail.com}
	
	\affiliation{Physics and Applied Mathematics Unit, Indian Statistical Institute, Kolkata 700108, India}

	\begin{abstract}
		\noindent
		In the evolutionary dynamics of a rock-paper-scissor (RPS) model, the effect of natural death plays a major role in determining the fate of the system. Coexistence, being an unstable fixed point of the model becomes very sensitive towards this parameter. In order to study the effect of mobility in such a system which has explicit dependence on mortality, we perform Monte Carlo simulation on a $2$-dimensional lattice having three cyclically-competing species. The spatio-temporal dynamics has been studied along with the two-site correlation function. Spatial distribution exhibits emergence of spiral patterns in the presence of mobility. It reveals that the joint effect of death rate and mobility (diffusion) leads to new  coexistence and extinction scenarios. 
		 		
	\end{abstract}	
	
	\maketitle
	
	\section{Introduction} 
	\label{intro}
	\noindent
	{A stable ecosystem consists of different coexisting species with naturally balanced intra- and inter-species interactions like offspring production, predation-prey, intrinsic natural death etc. A natural goal of any ecological system is to maintain the biodiversity such that any species can avoid its extinction.} To capture { the complexity in a stable ecological system, the principal idea behind evolutionary game dynamics is that the survival or success of species depends on others is very useful \cite{szabo2007evolutionary,perc2017statistical,szolnoki2014cyclic,pennisi2005determines, mobilia2016influence,traulsen2006stochastic}.}
	 A paradigmatic {model of} evolutionary game dynamics such as the rock-paper-scissor (RPS) model, where  cyclic dominance \cite{arunachalam2020rock, kerr2002local, park2019fitness, reichenbach2007mobility, hashimoto2018clustering}  determines the fate of each strategy, {often successfully mimics the emergence} of biodiversity in several natural systems. Colony formation and coexistence of several microbes \cite{ke2020effects, momeni2017lotka, kerr2002local,nahum2011evolution}, parasites \cite{cameron2009parasite, segura2013competition} etc. have been studied using this type of formalism. The stability of the {\it Uta stansburiana lizards} \cite{sinervo1996rock}, fermentation in the presence of oxygen and at high glucose concentrations \cite{pfeiffer2014evolutionary}  and diversity of coral-reef organisms \cite{jackson1975alleopathy} can  be explained with this type of cyclic game. 
	 {The study of evolutionary RPS model can be characterized \cite{sinervo1996rock} in two formalisms: Lotka-Volterra (LV) approach \cite{bacaer2011lotka, lotka1920analytical, volterra1926fluctuations} in which particular species densities are conserved and May-Leonard (ML) approach \cite{chi1998asymmetric} where vacant sites are introduced and thereby species densities can be varied.}
	 %
	 \par It has already been studied ({both theoretically and experimentally}) that parameters such as system size, mobility, interaction region, protection spillover, risk-averse
	 	hedgers affect the stability and evolution of the system in a significant manner \cite{kerr2002local,pennisi2005determines,frey2010evolutionary,muller2010community,szolnoki2015vortices, kelsic2015counteraction, guo2020novel}. In particular, demographic fluctuation breaks the coexistence leading to the extinction of certain species \cite{reichenbach2006coexistence,serrao2021rare,west2020fixation,berr2009zero}. Currently, {the role of individual natural death (mortality) has been examined on the cyclic interaction of three species \cite{bhattacharyya2020mortality}.} It is observed that suitable intrinsic death rate of a species may dominate the effect of the birth or predation rates. As a result, a tiny change in the death rate may lead the system towards extinction or make any one species dominate over the others. 
	\par On the other hand, mobile species i.e., species having some kind of exchange or hopping probability may {alter the erstwhile scenario of survival} \cite{kerr2002local,reichenbach2008self,kirkup2004antibiotic}. Depending on the strength of the mobility, several spatial structures such as { spiralling} pattern \cite{ reichenbach2008self,avelino2012junctions, avelino2018directional} may emerge in the  extended 2D systems. In certain cases, these spirals grow in size depending on the rate of exchange and after a critical value of the hopping rate, the system shows anomalous behaviour where only one species survives and other two go extinct. In particular, the size of spirals expands deliberately and it becomes larger than the system's size enabling any one of those species occupy the whole system \cite{reichenbach2007mobility}.
	\par In this backdrop, we revisit the effect of mobility in a cyclic RPS game dynamics where each species has its own intrinsic natural death rate. Using Monte Carlo (MC) simulations \cite{mooney1997monte, zio2013monte, szabo2007evolutionary}, we have explored the effects of mobility in a 2D RPS system within ML formalism where {each individual has a mortality rate i.e., all the species have a probability of death} irrespective of any other parameters like predation or reproduction. 
	{We have observed that mobility helps the species survive in an unfavourable environment where mortality rate is high.}  
	\par {The rest of the article is organized as follows. We describe the model, the dynamics and the method of simulation in Sec.~\ref{simulation}. The results are discussed in Sec.~\ref{results} and finally we conclude our remarks  in Sec.~\ref{conclusion}.}
	\begin{figure*}[ht!]
	\centering
	\includegraphics[width=1.1\textwidth]{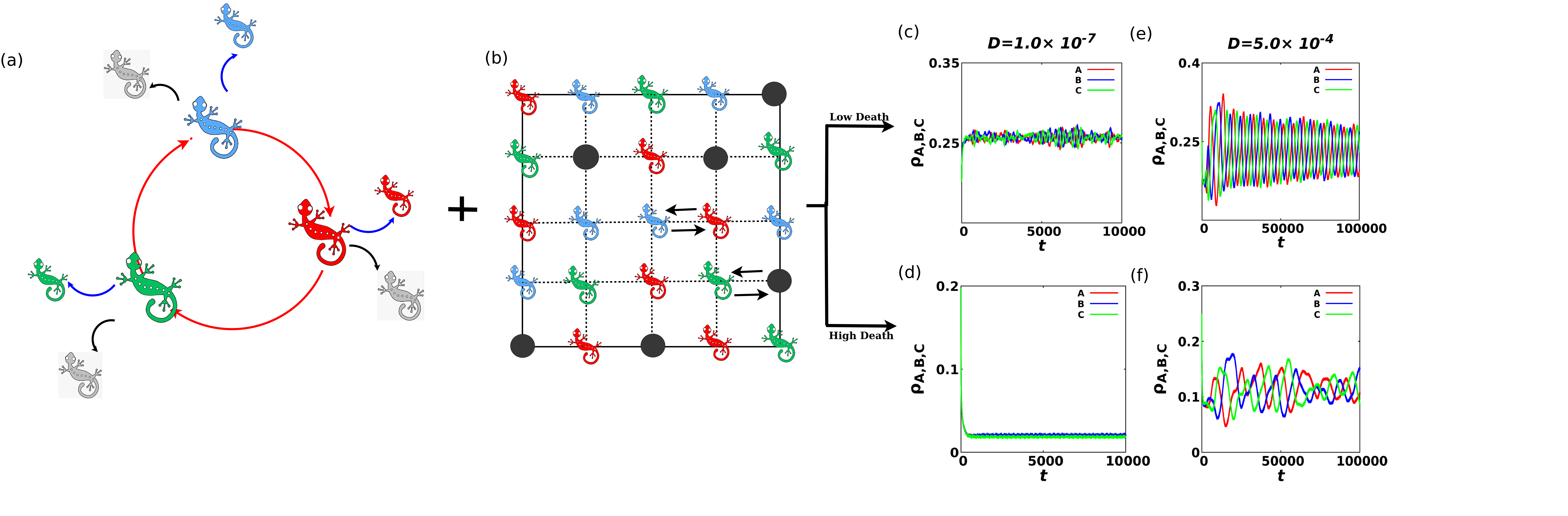}
	\caption{{\bf Schematic diagram of cyclic interaction and mobility in presence of natural death}. (a) Red arrows represent the cyclic predation between three species (red, blue, and green). Blue and black arrows reflect the reproduction  and natural death respectively. (b) Mobility or exchange between two nearest neighbours (one of them may be vacant site). (c) and (e) At low death rate ($d=1$), the densities of species remain finite for two mobility strength : smaller  $D=1\times 10^{-7}$  as well as in higher value $D=5\times 10^{-4}$. (d) At high death rate ($d=4$)  and for low diffusion, the density of each species is almost zero and no oscillatory behaviour is observed there. However  (f) in the presence of higher diffusion, the oscillation with higher amplitude appears. In all the cases, $1000\times 1000$ lattice has been considered with unnormalized predation and reproduction rates, $p=7$ and $r=7$ respectively for each of the species.}
	\label{fig1}
\end{figure*}

	\section{The formulation of the Model and  Monte Carlo simulation}
	\label{simulation}
	
	\noindent
	We consider three species in a 2D lattice where each lattice site can either have one species or be vacant. In the beginning of the simulation, the species and vacancies are randomly 
	{ distributed}. Then we perform MC simulation {governed by some interactions such as reproduction, predation, natural death and mobility all subjected to some predefined conditions}.
	\par {We denote the three species by $A,\;B$ and $C$ and the corresponding densities by $\rho_a,\; \rho_b$ and $\rho_c$ respectively. The vacancies are denoted by $V$ and their density, by $\rho_v$ where $\rho_a + \rho_b + \rho_c + \rho_v = 1$. In the cyclic domination process, the predations can be represented by the following equations.}   
	\bea
	A + B & \longrightarrow & A + V \;\;\; \mbox{with rate}\;\; p_a\nonumber \\
	B + C & \longrightarrow & B + V \;\;\; \mbox{with rate}\;\; p_b\nonumber \\
	C + A & \longrightarrow & C + V \;\;\; \mbox{with rate}\;\; p_c
	\label{predation}
	\eea 
	{where $p_{a,b,c}$ are the predation rates of $A,\;B$ and $C$ respectively. The reproductions can be expressed as}
   \bea
	A + V & \longrightarrow & A + A \;\;\; \mbox{with rate}\;\; r_a\nonumber \\
	B + V & \longrightarrow & B + B \;\;\; \mbox{with rate}\;\; r_b\nonumber \\
	C + V & \longrightarrow & C + C \;\;\; \mbox{with rate}\;\; r_c
	\label{reproduction}
	\eea
	{The rate of reproductions are denoted by $r_{a,b,c}$. Alongside, the natural death of a species can make a site vacant as well.}
	\be
	S \longrightarrow  V \;\;\; \mbox{with rate}\;\; d_{a,b,c} 
	\label{death}
	\ee
	{where $S$ represents any of the three species $A,\;B$ or $C$ and $V$ is the vacant site. The corresponding rate of death is $d_a,\;d_b$ or $d_c$. The cyclic predations, reproductions and deaths are schematically shown} in Fig.\ \ref{fig1}(a). Here red lizards (say, $A$) predate the greens ($B$), greens predate the blues ($C$) and blues predate the reds again. The blue arrows in Fig.\ \ref{fig1}(a) denote the birth of new species. Each lizard of  particular colour creates another one of the same colour (shown in smaller size). The black arrows indicate the natural death, and the grey lizards indicate the dead ones. Note that, we have considered identical predation, reproduction and death rate of each species: $p_a=p_b=p_c=p$, $r_a=r_b=r_c=r$ and $d_a=d_b=d_c=d$.

Finally, we have considered { nearest-neighbour pair exchange where}  the species can exchange their positions with that of the neighbouring species or vacancy governed by some probability:	
	\bea
	S + \phi(\neq S) & \longrightarrow & \phi(\neq S) + S \;\;\; \mbox{with rate}\;\; \epsilon
	\label{hop}
	\eea
Here $\phi$ can be either $A,\;B,\;C$ or $V$. Figure \ref{fig1}(b) describes this mobility process. {Here one blue and one red lizard in the  lattice   exchanged their position (marked by thick black arrow). In a similar way,  a lizard (see the green one) may hop to a vacant site (described as black filled dots). Note that, both swaps (between species-species or species-vacant) do not occur simultaneously but in two different MC step}. We consider only nearest neighbour interactions and assume periodic boundary condition while doing the simulation. 
The simulation starts with a random initial configuration. At each MC step one lattice site is considered randomly. {Then another site is chosen out of the four nearest neighbours of the first site and possible operations (predation, reproduction, exchange) are made between two sites with specified probabilities. Apart from this, the chosen first site is converted to a vacant one 
	 { with death rate $d$}. All the probabilities are normalized by the rates $r,p,d$ and $ \epsilon$. This entire process is repeated until an equilibration is reached.} {Numerical simulations have been performed on  $1000 \times 1000$ lattice,
and total number of required MC steps  varied from $10^4$ to $10^6$ depending on the  rate of mobility.} 		
The exchange process in Eq.~(\ref{hop}) leads to an effective diffusion of the individual species governed by the macroscopic diffusion constant $D$. {In our case of a 2-dimensional system, we connect this diffusion constant $D$ with the exchange rate $\epsilon$ through system size by the relation, $\epsilon=2 N^2 D $. \cite{reichenbach2007noise} }

		\begin{figure}
		\includegraphics[width=0.5\textwidth, height=4.5cm]{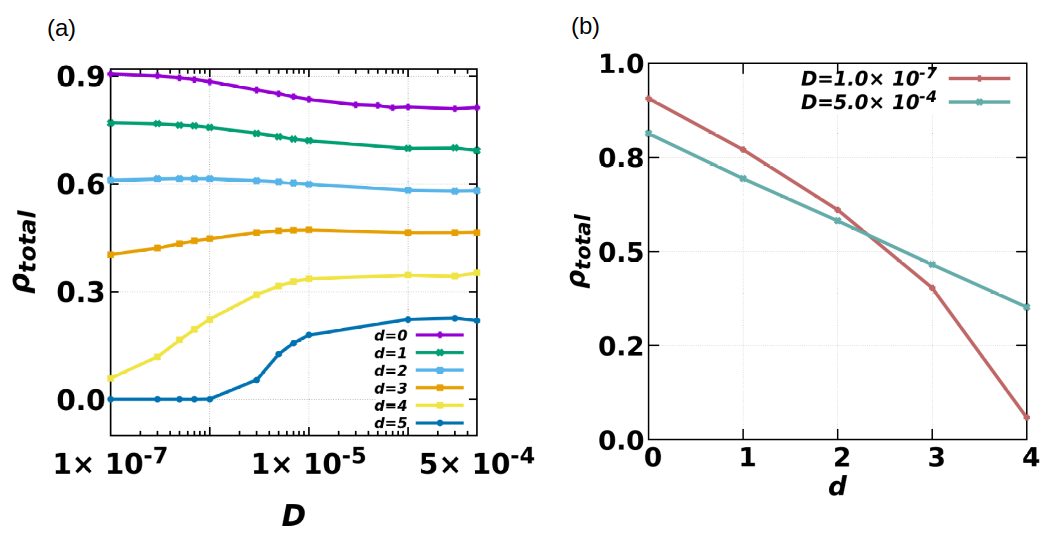}
		\caption{{\bf Role of mobility and death rate on species coexistence.} (a) Total species densities are plotted against mobility strength ($D$) for six constant death rates. We observe that at lower death rates as the mobility increases the total species density decreases slowly. At high death rate and low mobility the species density is very low (gets extinct for $d=5$). However, as the mobility increases species density increases (``revives'' even for $d=5$) and coexists. 
			(b) Total species density is plotted for two extreme mobility constants ($ D=1.0\times 10^{7}$ and $ D=5.0\times 10^{-4}$ ) with increasing death rate. With increasing death rate, the total species density decreases, but the slope for these two cases are different. Lower mobility leads to faster extinction but higher mobility decreases the slope and leads to non-extinction of the species for a wider range of death rate.}
		\label{fig2-speciesdensity}
	\end{figure}
\section{Simulation results}
\label{results}
{It has already been established \cite{bhattacharyya2020mortality} that the system has sharp dependence on natural death as compared to predation and reproduction. For a certain set of parameter values \cite{bhattacharyya2020mortality} the system also exhibits coexistence of all the species resulting from an unstable fixed point having all non-zero densities. Now, with the incorporation of diffusion (or mobility) the fluctuation expectedly increases in the dynamics. 
Figure \ref{fig1}(c-f) reports the density of the species with respect to time for two different death rates and two different mobility rates. {In lower (higher) death rate, the density of each species shows a stable  oscillation with high (small) amplitude.}  However the fluctuations are noted to be increased considerably when the mobility rate increases from $1.0\times10^{-7}$ [Fig.~\ref{fig1}(c) \& (d)] to $5.0\times10^{-4}$ [Fig.~\ref{fig1}(e) \& (f)].}

\begin{figure*}
	\includegraphics[width=16.9cm, height=13cm]{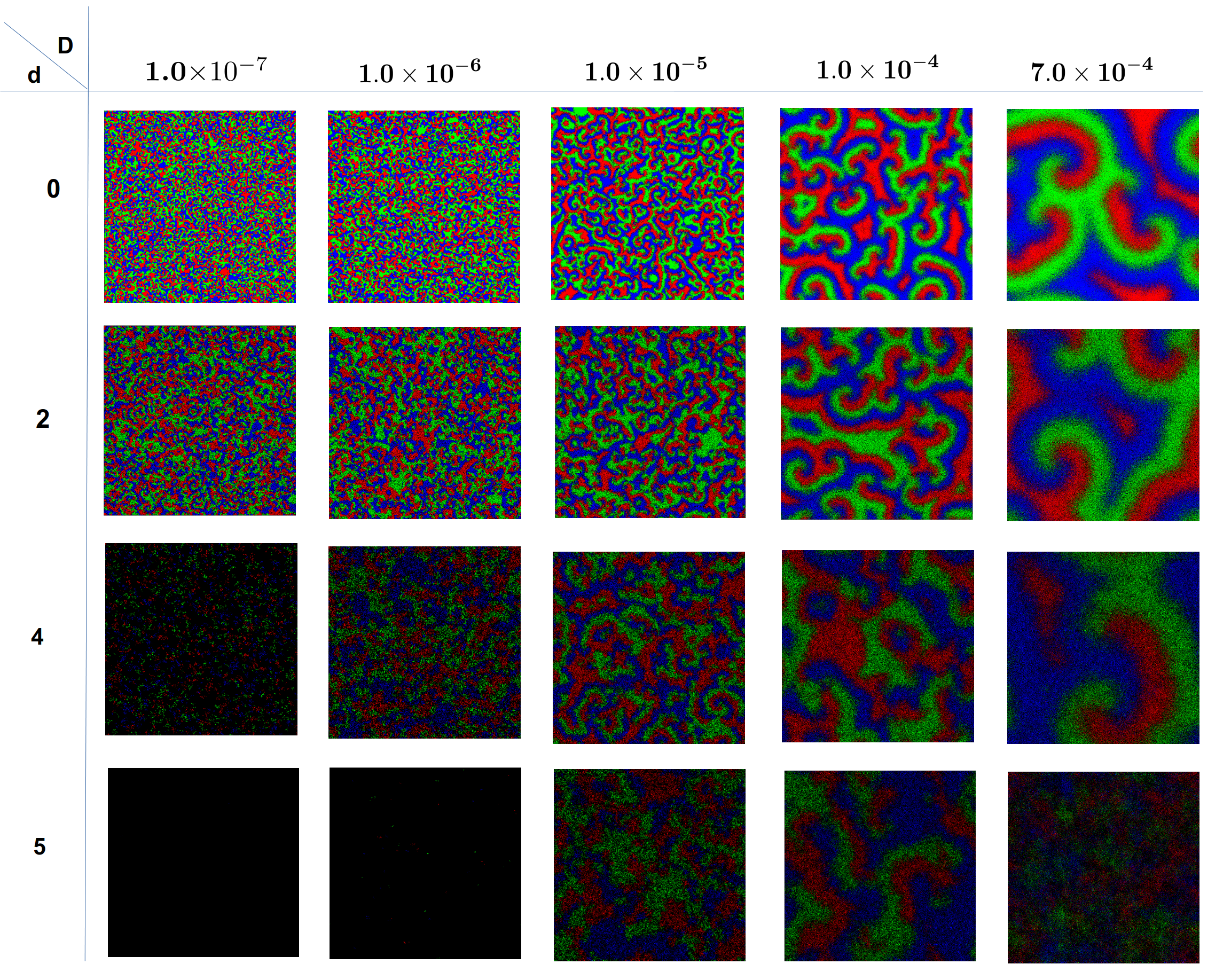}
	\caption{{\bf Spatial plot of species density for different death rate and mobility strength.} Distribution of the species throughout the lattice is shown by colours : Red, Green and Blue for $A$, $B$, and $C$ respectively. Black dots are  used for blank space. Diffusion strength is increased from $10^{-7}$ to $7\times 10^{-4}$ (from left to right). Natural death rate is increased from zero to $5$ (upper to bottom). }
	\label{fig3-spiral}
\end{figure*}
Increase of mobility rate is observed to affect the average densities of the species significantly in the regime of high as well as low death rate. When death rate is high the average species densities remain very low leaving most of the lattice sites vacant for low mobility (e.g see Fig.\ \ref{fig1}(d) for $d=4$). 
However, an increasing value of diffusion 
 { enhances} the species densities for the same rate of death. We can see in Fig.~\ref{fig1}(f) that the average density of the species has been raised as mobility becomes $5.0\times10^{-4}$. The situation becomes opposite when we are in the low death regime. Increasing mobility lowers the species densities there. 
  { This feature indicates that a coexistence regime may arise under high death rates within a critical boundary of mobility } 
{ We explore the combined effect of death rates and mobility for finding the coexistence as well as the the extinction regime. }.

\par To delve deeper, we have calculated the {total density of species ($1 - \rho_v$)} in the lattice as a function of mobility strength for different death rates. {The time-averaged total density is calculated following the definition
\be \rho_{\rm total} = \dfrac{1}{t_f - t_i}\sum\limits_{t_i}^{t_f} (\rho_a + \rho_b + \rho_c)\ee
where both $t_i$ and $t_f$ are in the equilibrated region of the MC simulation. Fig.\ \ref{fig2-speciesdensity}(a) shows the plots of $\rho_{\rm total}$ against mobility rate for six different death rates.} At lower mobility ($D\sim 10^{-7}$) and low death rates ($d=0$ and $d=1$ i.e., purple and green lines), the three species collectively occupy around $90\%$ and $80\%$ space of the whole lattice respectively. When the mobility rate increases, $\rho_{\rm total}$ decreases continuously and saturates at certain values ($\sim 0.8$ for purple and $\sim 0.7$ for green line). The total density remains almost unaltered with varying mobility for moderate death rates ($d\sim 2$). On the other hand, for species with higher death rates ($d\gtrsim 3$), an increase in $\rho_{\rm total}$ is observed with increasing mobility rate. For very high death rate ($d\sim 5$), { coexistence of species is observed. By contrast,  the system shows extinction at  low diffusion.}
In Figure  \ref{fig2-speciesdensity} (b), we have plotted the total species density as a function of  death rates for two extreme mobility strengths: ($D=1.0\times10^{-7}$ and $D=5.0\times10^{-4}$). We observe a monotonic decrease of $\rho_{\rm total}$ with the death rate $d$. As the Figure shows that  the lower mobility rate (magenta line) has faster decay in  $\rho_{\rm total}$ than the higher mobility case (green line). 
{ This leads to the fact that the system with higher mortality can overcome extinction  for certain range of higher mobility value.}
 Additionally, we observe a crossover of the two  lines where the total density is around $0.55$. 
This intersection point gives a certain value of death rate and species density where two extreme mobility rates lead to an identical population density. 
\\
For further analysis, we have plotted the species density spatially in a  $1000\times 1000$ lattice (Fig.\  \ref{fig3-spiral}). { Each of the figure is plotted  when the system had already reached  stable oscillation well ahead.} In the first row, where no natural death rate is considered, the size of the spirals increase from low mobility to higher mobility ($D=1.0\times 10^{-7}$ to $D=1\times 10^{-4}$ ). {If we increase the mobility rate further ($D\geq7.0\times 10^{-4}$), then the size of one spiral would be inflated so much that it would occupy the entire lattice leading to the situation of single species survival {(not shown here)}.} 
This is consistent with the previously reported results \cite{reichenbach2007mobility}. Now, if we introduce the natural death of each species, the size of the spirals grow more rapidly  
(second row, from left to right). In the third row, at death rate $d=4$ and at mobility strength $D=1\times 10^{-7}$ a large fraction of the lattice sites become vacant. Thus a large collection of black dots appear in the spatial plot. However, if we increase $D$, the large spirals reappear in the lattice. 
{Panels of 2 bottom rows, for $d=4$ and $d=5$, become less dark as $D$ is increased (left to right). }
{Thus, the mobility has capability of 
	{ countering}  the effect of death rate of the species. The mathematical analysis of the emergence of  spiral patterns  \cite{ipsen2000amplitude,ipsen2000amplitudee,gong2003antispiral,mondal2019diffusion,szczesny2014characterization,mondal2021spatiotemporal} will be explored in future. \\
\begin{figure*}
	
	 \includegraphics[width=18 cm,height=5.6 cm]{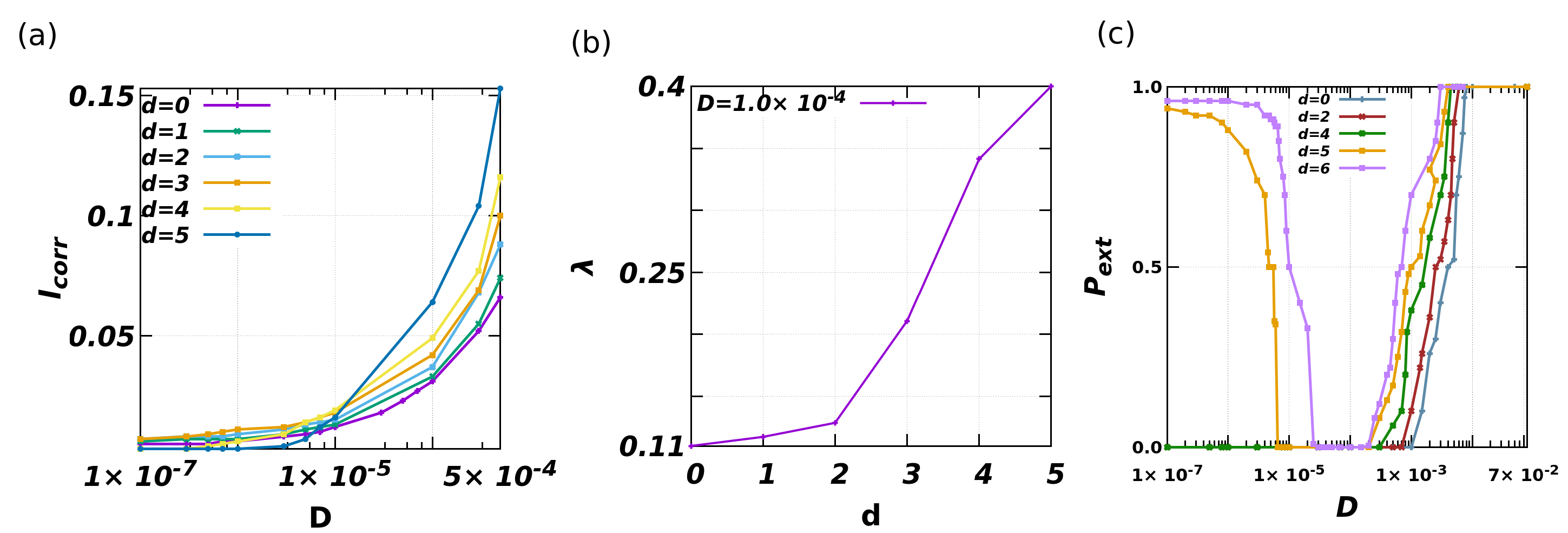} 

	\caption{ 
		(a) Correlation lengths ($l_{\rm corr}$) are plotted against different mobility strengths ($D$). {At  higher diffusion}, the length is enhanced. This is consistent with previous results \cite{reichenbach2007mobility}. If the death rate is increased, then the length is also sharply increased. In intermediate mobility strength ($D\sim 10^{-5}$), the species density ``revives'' (species coexistence) for higher death rate thus correlation becomes non zero (see the blue and yellow line). 
	{  (b) 
		The wavelength ($\lambda$) of the correlation function as a function of death ($d$) for a particular value of mobility ($D=1.0\times10^{-4}$). The wavelength is increasing as we increase the death value. This indicates the spirals increase in size with increasing death rate. } 
	(c)  Extinction probability as a function of mobility $D$. Five  death rates have been chosen. For higher death rates ($d=5,6$), the extinction probability is high for low ($D\sim 10^{-7}$) and high ($D\gtrsim 10^{-3}$) diffusion, and goes to zero in the intermediate values of $D \sim 10^{-4}$.}
	\label{fig4-corr}
\end{figure*}

\begin{figure*}
	 \includegraphics[width= \textwidth,height=5  cm]{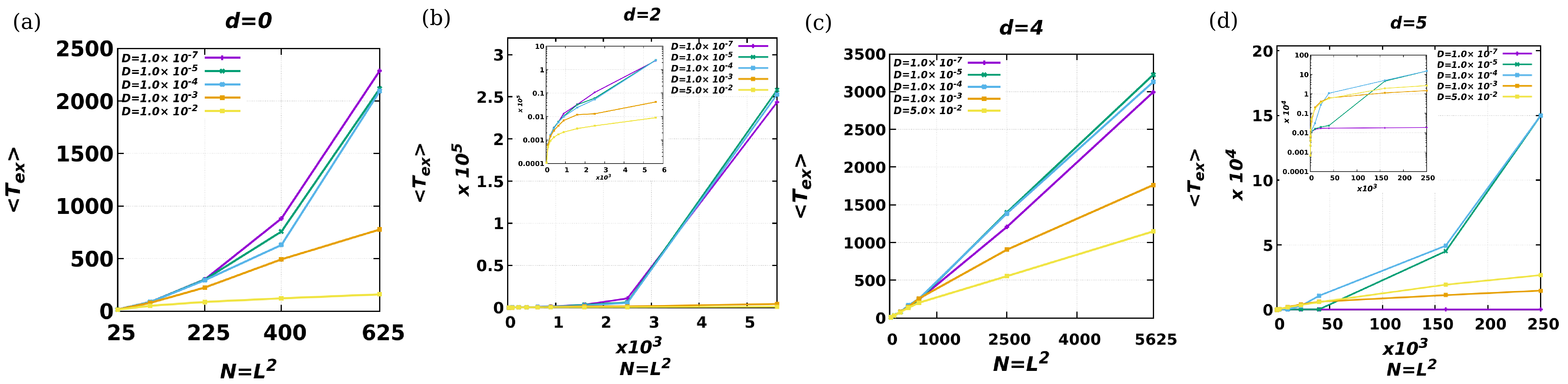} 
	\caption{{\bf Effect of lattice size on average extinction time at different death rates and mobility parameters:} $\langle T_{\rm ex}\rangle$ is plotted against lattice size, $N\;(=L^2)$ for different death rates ($d$) and mobility ($D$). (a-b) Diffusion dominates for low death rates as it lowers the extinction time. (c-d) For high death rates both low and high diffusivity show lower extinction times whereas $\langle T_{\rm ex}\rangle$ is raised for an intermediate range of diffusion.  Note that in (d) data for  $d=5, N=500^2$ 
	and $D=10^{-4}$ and $D=10^{-5}$
	are given as eye guides since coexistence is not lost at time $1.5\times 10^5$ (hence, in these cases $\langle T_{\rm ex}\rangle> 1.5\times 10^5$), see text.
	 Insets of (b) and (d): $\langle T_{\rm ex}\rangle$ vs. $N$ on lin-log scale.
	 In this work, in one time step there are $N$ elementary update moves 
	corresponding  to $N=L^2$ Monte Carlo steps.
	}
	\label{Fig-T_exvs_N}
\end{figure*}
{The patterns can also be realized from the study of the spatial correlation}. The {two-site} correlation function is defined by:
	\begin{eqnarray}
g_{s_i s_j}(\mathbf{r}-\mathbf{r^\prime} , t) &\equiv& \langle s_i(\mathbf{r},t) s_j(\mathbf{r^\prime},t) \rangle -\langle s_i(\mathbf{r},t) \rangle  \langle  s_j(\mathbf{r^\prime},t) \rangle
\end{eqnarray}
\noindent
Here $s_{i,j}(\mathbf{r},t)$ ($i,j\in a,b,c$) represents  a species at the position $\mathbf{r}$ in the $2D$ lattice at a particular time $t$ \cite{reichenbach2007noise}. {We study the correlation of species $A$ ($g_{s_a s_a}$) and it shows damped oscillation with respect to distance in case of the spiral patterns as observed in earlier studies \cite{reichenbach2007noise,he2011coexistence}.} The correlation length ($l_{\rm corr}$) is thereby determined by the length at which the correlation drops $1/e ^{th}$ of the maximum value \cite{reichenbach2008self}. {{ As expected \cite{reichenbach2007mobility},} all the plots show that the correlation length increases with increasing mobility indicating the size of a single patch in the spatial distribution getting bigger.} 
{At lower diffusion, the correlation length for high death rate is almost nil mainly due to the extinction of species at that regime (Fig.\ \ref{fig4-corr}(a)). At higher mobility ($D\sim 10^{-5}$) 
 { a new coexistence regime appears } (the blue and yellow line, also see the third and fourth row of the Fig.\ \ref{fig3-spiral}) significantly boosts up the correlation length curve jointly with the inflated spiral size. 
{ The size of the spiral increases with the death rate as well. This is indicated by Fig.~\ref{fig4-corr}(b) where we have plotted the wavelength ($\lambda$) of the correlation function ($g_{s_a s_a}$) against death rates.}

Interestingly in the regime of high diffusion, the growth of the correlation length, i.e. the size of the patch is elevated and the significance is that the species are prone to colony formation more when their mortality rates are high. This feature will be explored in more details in future. 
\par {
{ 	We have also calculated the extinction probability ($P_{\rm ext}$, Fig.\ \ref{fig4-corr}(c)) against mobility strength for different death rates. { Here $P_{\rm ext}$ is the probability of extinction of one or more species \cite{reichenbach2007mobility}, numerically calculated over a large number of realizations for a lattice size of  $1000\times 1000$ and for $10^6$  MC steps. }  It is already established that \cite{reichenbach2007mobility}    an abrupt transition occurs from coexistence to extinction (size $N\rightarrow \infty$) above a certain critical $D$  in the case of  $d=0$. In finite systems, there are finite-size effects as shown in  Fig.\ \ref{fig4-corr}(b) (blue, red and green). }	
For higher mobility ($D>10^{-3}$), the coexistence is completely lost, as the patch of one species becomes big enough to cover the entire space. The results are almost same for death rates $d=2$ (brown) and $d=4$ (green). In the presence of death, the system reaches extinction  
{ at smaller diffusion} compared to zero mortality rate.
{ However, an interesting feature is observed for higher mortality rates. For large $d$ ($d=5$ (yellow) and $d=6$ (violet)),  new extinction and coexistence regimes appear that is characterized by two critical values of $D$:  $D_{c1}$ and $D_{c2}$.
For $D<D_{c1}$  and  $D>D_{c2}$ one species takes over, coexistence is lost after a time of { order $N$}, (effect of lattice size  is explored  in Fig.\ \ref{Fig-T_exvs_N}).
 For $D_{c1}<D<D_{c2}$  long-lived coexistence of all species (metastable state) occurs for a time that grows exponentially with $N$ (Fig.\ \ref{Fig-T_exvs_N}). For instance, at $d=5$, there is only one surviving species for $D \lesssim 10^{-5}$ and  $D \gtrsim 10^{-3}$ and species coexist for $10^{-5}<D<10^{-3}$. Here, the joint effect of high death rate and mobility leads to new extinction/ coexistence regimes or scenarios.  Below we provide an intuitive interpretation of these findings.
   Note that we have confirmed this scenario and the values of $P_{\rm ext}$, $D_{c1}$ and $D_{c2}$, for larger systems in order to rule out the influence of finite size effects.
 \par 
 When $d$ is large, there are numerous spontaneous death events leading to a lot of empty spaces. Moreover, when there is enough mobility (high $D$), individuals visit a large fraction of the lattice, and hence can efficiently find available  empty sites
 where to  reproduce before being killed. In this sense high mobility rate $D$ counters high death rates, and allows species to coexist under extreme conditions (high $d$) provided that they are sufficiently mobile (high enough $D$). Of course, if $D$ is too large spiralling patterns just outgrow the system as in \cite{reichenbach2007mobility}. When $d$ is small, we also recover essentially the same extinction/coexistence scenario as in \cite{reichenbach2007mobility}.
 
%
{	Novel phenomenology hence arises when $d$ is large and $D$ is sufficiently large (but not too large), mobility counters the effect of killing and allows species to coexist for a very long time (see below and Fig.~\ref{Fig-T_exvs_N}). 
	In the opposite limit, when $D$ is small ($d$ being large), individuals are not mobile enough to escape spontaneous death and species coexistence ceases after a finite time.
}
} 


}	
 	
\par  Finally we have investigated the effect of finite size in the dynamics. We have chosen four death rates: $d=0,2,4,~{\rm and} ~5$. The average extinction time $\langle T_{\rm ex}\rangle$ of the species is plotted with the system size $N$ (Fig.\ \ref{Fig-T_exvs_N}(a-d)).   The 
	extinction time {($T_{\rm ex}$)} is measured when one of the species goes to extinction. For each death rate and diffusion, {the average extinction time,} $\langle T_{\rm ex}\rangle$ is calculated for $100$ realizations (1 time step $=N$ Monte Carlo steps $=N$ update moves). The mean extinction time
$\langle T_{\rm ex}\rangle$ increases with increasing system size and the nature of the curve depends on the strength of mobility as observed previously in absence of any death rate \cite{he2011coexistence}. The death rate also has influence on $\langle T_{\rm ex}\rangle$. In the case of low (or no) death rates (Fig.\ \ref{Fig-T_exvs_N}(a-b)), low mobility is observed to lower the extinction time.
{ For instance,  $d=2$, $P_{\rm ext}\approx 1$ at $D>10^{-3}$ and $P_{\rm ext}\approx 0$ for $D<10^{-3}$. This suggests that for $d=2$ and $D<10^{-3}$ the coexistence is metastable and extinction occurs in a  time that appears to be scaling exponentially with $N$, as suggested by Fig.~\ref{Fig-T_exvs_N}(b)(See inset of Fig.~\ref{Fig-T_exvs_N}(b)). On the other hand, when $D>10^{-3}$ the systems settles rapidly in a regime where two species have gone extinct and $\langle T_{\rm ex}\rangle$ seems to be approximately linear in $N$ when $N\gg 1$, as in Ref.~\cite{he2011coexistence}.
In the novel coexistence/extinction scenario found here, 
coexistence is metastable at large $d$ and sufficiently large $D$, e.g. for $d=5$ and $D=10^{-5}-10^{-4}$, a regime in which  $P_{\rm ext}\approx 0$ and all species coexist for a  time that appears to be scaling exponentially with $N$, see Fig.~\ref{Fig-T_exvs_N}(d) and its inset. 
Note however that the  data point in  Fig.~\ref{Fig-T_exvs_N}(d), for $N=500\times 500$, with $d=5$, 
 $D=10^{-4}$ and $D=10^{-5}$, simulations were performed for $1.5\times 10^{5}$ time steps after which species coexistence persists, from which we infer that in these cases $\langle T_{\rm ex}\rangle>1.5\times 10^{5}$ which appears to be compatible with $\langle T_{\rm ex}\rangle$ growing exponentially with $N$ when $N\gg 1$. 
However, at fixed large $d$, when $D$ is either very high or too low, e.g. for $D=10^{-7}$ and $D\gtrsim 10^{-3}$ in Fig.~\ref{Fig-T_exvs_N}(d), $P_{\rm ext}\approx 1$. In these regimes of very low/high $D$, two species go extinct on a much shorter time scale, with $\langle T_{\rm ex}\rangle$ that appears  
to grow approximately linearly in $N$ when $N\gg 1$.

}



%

\section{conclusion}
\label{conclusion}	
	{ We study the role of mobility in the spatiotemporal behavior of a 3-species ecosystem with cyclically dominating interaction. We also incorporate a rate of natural death which represent the finite lifetime of the living system. The ecosystem is mapped into a 2-dimensional lattice on which the RPS dynamics is studied in ML formalism through Monte Carlo simulation.
We mainly concentrate in the parameter region where the system exhibits coexistence. The natural death rate has already been proven to affect the coexistence significantly. Our present study reveals that high mobility rate overshadows the act of natural death.} 
We have demonstrated how the time-averaged total density of the competing species changes with the rate of death as well as the rate of diffusion.
{In addition, 
{the spatiotemporal analysis followed by the two-site correlation length suggests that the size of the patches formed by individual species increases with both $D$ and $d$}. Interestingly,   a novel extinction and coexistence regimes appears under high death rates and characterized by two critical mobility rates between which species coexist.}	
 \par In fact high death rate leads to an increase of vacant sites leading to more opportunity for random mixing than when $d=0$. Again the increase of correlation with the death rate suggests that increasing $d$ at fixed $D$ would lead to larger patches of activities, but with blurrier shapes and interfaces due to more random mixing.}
{The results therefore suggests that an ecosystem with its constituent species being highly mobile can evade possible extinction caused by increased mortality rate.}
It has recently been shown that the  introduction of pestilent species attacking over a single species may jeopardize the stability of the coexistence \cite{bazeia2021effects}. We would like to explore this feature in presence of natural death and mobility.  Also, de Oliveira {\it et al} \cite{de2021mobility} have checked the effect of mobility in a lattice of living organisms \cite{drescher2014solutions}. Instead of global restrictions on reproduction and movement, these authors considered local restrictions and showed that these were able to generate multicluster states. In future we would like to explore the effect of breaking the directional symmetry of the movement on the appearance of the    novel coexistence and the extinction scenario.
\par A diffusive system system allows inter-species exchange which eventually promotes colony formation among the species. However the influence of mobility in the effect of death rate is counterintuitive because mobility and mortality as incorporated in this system, are apparently two unconnected phenomena.

	\section{Acknowledgements} CH is supported by INSPIRE-Faculty grant (Code: IFA17-PH193). \\
	
	\bibliography{bibliography_main}

\begin{thebibliography}{50}%
\makeatletter
\providecommand \@ifxundefined [1]{%
 \@ifx{#1\undefined}
}%
\providecommand \@ifnum [1]{%
 \ifnum #1\expandafter \@firstoftwo
 \else \expandafter \@secondoftwo
 \fi
}%
\providecommand \@ifx [1]{%
 \ifx #1\expandafter \@firstoftwo
 \else \expandafter \@secondoftwo
 \fi
}%
\providecommand \natexlab [1]{#1}%
\providecommand \enquote  [1]{``#1''}%
\providecommand \bibnamefont  [1]{#1}%
\providecommand \bibfnamefont [1]{#1}%
\providecommand \citenamefont [1]{#1}%
\providecommand \href@noop [0]{\@secondoftwo}%
\providecommand \href [0]{\begingroup \@sanitize@url \@href}%
\providecommand \@href[1]{\@@startlink{#1}\@@href}%
\providecommand \@@href[1]{\endgroup#1\@@endlink}%
\providecommand \@sanitize@url [0]{\catcode `\\12\catcode `\$12\catcode
  `\&12\catcode `\#12\catcode `\^12\catcode `\_12\catcode `\%12\relax}%
\providecommand \@@startlink[1]{}%
\providecommand \@@endlink[0]{}%
\providecommand \url  [0]{\begingroup\@sanitize@url \@url }%
\providecommand \@url [1]{\endgroup\@href {#1}{\urlprefix }}%
\providecommand \urlprefix  [0]{URL }%
\providecommand \Eprint [0]{\href }%
\providecommand \doibase [0]{http://dx.doi.org/}%
\providecommand \selectlanguage [0]{\@gobble}%
\providecommand \bibinfo  [0]{\@secondoftwo}%
\providecommand \bibfield  [0]{\@secondoftwo}%
\providecommand \translation [1]{[#1]}%
\providecommand \BibitemOpen [0]{}%
\providecommand \bibitemStop [0]{}%
\providecommand \bibitemNoStop [0]{.\EOS\space}%
\providecommand \EOS [0]{\spacefactor3000\relax}%
\providecommand \BibitemShut  [1]{\csname bibitem#1\endcsname}%
\let\auto@bib@innerbib\@empty
\bibitem [{\citenamefont {Szab{\'o}}\ and\ \citenamefont
  {Fath}(2007)}]{szabo2007evolutionary}%
  \BibitemOpen
  \bibfield  {author} {\bibinfo {author} {\bibfnamefont {G.}~\bibnamefont
  {Szab{\'o}}}\ and\ \bibinfo {author} {\bibfnamefont {G.}~\bibnamefont
  {Fath}},\ }\href@noop {} {\bibfield  {journal} {\bibinfo  {journal} {Physics
  reports}\ }\textbf {\bibinfo {volume} {446}},\ \bibinfo {pages} {97}
  (\bibinfo {year} {2007})}\BibitemShut {NoStop}%
\bibitem [{\citenamefont {Perc}\ \emph {et~al.}(2017)\citenamefont {Perc},
  \citenamefont {Jordan}, \citenamefont {Rand}, \citenamefont {Wang},
  \citenamefont {Boccaletti},\ and\ \citenamefont
  {Szolnoki}}]{perc2017statistical}%
  \BibitemOpen
  \bibfield  {author} {\bibinfo {author} {\bibfnamefont {M.}~\bibnamefont
  {Perc}}, \bibinfo {author} {\bibfnamefont {J.~J.}\ \bibnamefont {Jordan}},
  \bibinfo {author} {\bibfnamefont {D.~G.}\ \bibnamefont {Rand}}, \bibinfo
  {author} {\bibfnamefont {Z.}~\bibnamefont {Wang}}, \bibinfo {author}
  {\bibfnamefont {S.}~\bibnamefont {Boccaletti}}, \ and\ \bibinfo {author}
  {\bibfnamefont {A.}~\bibnamefont {Szolnoki}},\ }\href@noop {} {\bibfield
  {journal} {\bibinfo  {journal} {Physics Reports}\ }\textbf {\bibinfo {volume}
  {687}},\ \bibinfo {pages} {1} (\bibinfo {year} {2017})}\BibitemShut {NoStop}%
\bibitem [{\citenamefont {Szolnoki}\ \emph {et~al.}(2014)\citenamefont
  {Szolnoki}, \citenamefont {Mobilia}, \citenamefont {Jiang}, \citenamefont
  {Szczesny}, \citenamefont {Rucklidge},\ and\ \citenamefont
  {Perc}}]{szolnoki2014cyclic}%
  \BibitemOpen
  \bibfield  {author} {\bibinfo {author} {\bibfnamefont {A.}~\bibnamefont
  {Szolnoki}}, \bibinfo {author} {\bibfnamefont {M.}~\bibnamefont {Mobilia}},
  \bibinfo {author} {\bibfnamefont {L.-L.}\ \bibnamefont {Jiang}}, \bibinfo
  {author} {\bibfnamefont {B.}~\bibnamefont {Szczesny}}, \bibinfo {author}
  {\bibfnamefont {A.~M.}\ \bibnamefont {Rucklidge}}, \ and\ \bibinfo {author}
  {\bibfnamefont {M.}~\bibnamefont {Perc}},\ }\href@noop {} {\bibfield
  {journal} {\bibinfo  {journal} {Journal of the Royal Society Interface}\
  }\textbf {\bibinfo {volume} {11}},\ \bibinfo {pages} {20140735} (\bibinfo
  {year} {2014})}\BibitemShut {NoStop}%
\bibitem [{\citenamefont {Pennisi}(2005)}]{pennisi2005determines}%
  \BibitemOpen
  \bibfield  {author} {\bibinfo {author} {\bibfnamefont {E.}~\bibnamefont
  {Pennisi}},\ }\href@noop {} {\bibfield  {journal} {\bibinfo  {journal}
  {Science}\ }\textbf {\bibinfo {volume} {309}},\ \bibinfo {pages} {90}
  (\bibinfo {year} {2005})}\BibitemShut {NoStop}%
\bibitem [{\citenamefont {Mobilia}\ \emph {et~al.}(2016)\citenamefont
  {Mobilia}, \citenamefont {Rucklidge},\ and\ \citenamefont
  {Szczesny}}]{mobilia2016influence}%
  \BibitemOpen
  \bibfield  {author} {\bibinfo {author} {\bibfnamefont {M.}~\bibnamefont
  {Mobilia}}, \bibinfo {author} {\bibfnamefont {A.~M.}\ \bibnamefont
  {Rucklidge}}, \ and\ \bibinfo {author} {\bibfnamefont {B.}~\bibnamefont
  {Szczesny}},\ }\href@noop {} {\bibfield  {journal} {\bibinfo  {journal}
  {Games}\ }\textbf {\bibinfo {volume} {7}},\ \bibinfo {pages} {24} (\bibinfo
  {year} {2016})}\BibitemShut {NoStop}%
\bibitem [{\citenamefont {Traulsen}\ \emph {et~al.}(2006)\citenamefont
  {Traulsen}, \citenamefont {Nowak},\ and\ \citenamefont
  {Pacheco}}]{traulsen2006stochastic}%
  \BibitemOpen
  \bibfield  {author} {\bibinfo {author} {\bibfnamefont {A.}~\bibnamefont
  {Traulsen}}, \bibinfo {author} {\bibfnamefont {M.~A.}\ \bibnamefont {Nowak}},
  \ and\ \bibinfo {author} {\bibfnamefont {J.~M.}\ \bibnamefont {Pacheco}},\
  }\href@noop {} {\bibfield  {journal} {\bibinfo  {journal} {Physical Review
  E}\ }\textbf {\bibinfo {volume} {74}},\ \bibinfo {pages} {011909} (\bibinfo
  {year} {2006})}\BibitemShut {NoStop}%
\bibitem [{\citenamefont {Arunachalam}(2020)}]{arunachalam2020rock}%
  \BibitemOpen
  \bibfield  {author} {\bibinfo {author} {\bibfnamefont {A.}~\bibnamefont
  {Arunachalam}},\ }\href@noop {} {\bibfield  {journal} {\bibinfo  {journal}
  {Wireless Personal Communications}\ }\textbf {\bibinfo {volume} {113}},\
  \bibinfo {pages} {1315} (\bibinfo {year} {2020})}\BibitemShut {NoStop}%
\bibitem [{\citenamefont {Kerr}\ \emph {et~al.}(2002)\citenamefont {Kerr},
  \citenamefont {Riley}, \citenamefont {Feldman},\ and\ \citenamefont
  {Bohannan}}]{kerr2002local}%
  \BibitemOpen
  \bibfield  {author} {\bibinfo {author} {\bibfnamefont {B.}~\bibnamefont
  {Kerr}}, \bibinfo {author} {\bibfnamefont {M.~A.}\ \bibnamefont {Riley}},
  \bibinfo {author} {\bibfnamefont {M.~W.}\ \bibnamefont {Feldman}}, \ and\
  \bibinfo {author} {\bibfnamefont {B.~J.}\ \bibnamefont {Bohannan}},\
  }\href@noop {} {\bibfield  {journal} {\bibinfo  {journal} {Nature}\ }\textbf
  {\bibinfo {volume} {418}},\ \bibinfo {pages} {171} (\bibinfo {year}
  {2002})}\BibitemShut {NoStop}%
\bibitem [{\citenamefont {Park}(2019)}]{park2019fitness}%
  \BibitemOpen
  \bibfield  {author} {\bibinfo {author} {\bibfnamefont {J.}~\bibnamefont
  {Park}},\ }\href@noop {} {\bibfield  {journal} {\bibinfo  {journal} {EPL
  (Europhysics Letters)}\ }\textbf {\bibinfo {volume} {126}},\ \bibinfo {pages}
  {38004} (\bibinfo {year} {2019})}\BibitemShut {NoStop}%
\bibitem [{\citenamefont {Reichenbach}\ \emph
  {et~al.}(2007{\natexlab{a}})\citenamefont {Reichenbach}, \citenamefont
  {Mobilia},\ and\ \citenamefont {Frey}}]{reichenbach2007mobility}%
  \BibitemOpen
  \bibfield  {author} {\bibinfo {author} {\bibfnamefont {T.}~\bibnamefont
  {Reichenbach}}, \bibinfo {author} {\bibfnamefont {M.}~\bibnamefont
  {Mobilia}}, \ and\ \bibinfo {author} {\bibfnamefont {E.}~\bibnamefont
  {Frey}},\ }\href@noop {} {\bibfield  {journal} {\bibinfo  {journal} {Nature}\
  }\textbf {\bibinfo {volume} {448}},\ \bibinfo {pages} {1046} (\bibinfo {year}
  {2007}{\natexlab{a}})}\BibitemShut {NoStop}%
\bibitem [{\citenamefont {Hashimoto}\ \emph {et~al.}(2018)\citenamefont
  {Hashimoto}, \citenamefont {Sato}, \citenamefont {Ichinose}, \citenamefont
  {Miyazaki},\ and\ \citenamefont {Tainaka}}]{hashimoto2018clustering}%
  \BibitemOpen
  \bibfield  {author} {\bibinfo {author} {\bibfnamefont {T.}~\bibnamefont
  {Hashimoto}}, \bibinfo {author} {\bibfnamefont {K.}~\bibnamefont {Sato}},
  \bibinfo {author} {\bibfnamefont {G.}~\bibnamefont {Ichinose}}, \bibinfo
  {author} {\bibfnamefont {R.}~\bibnamefont {Miyazaki}}, \ and\ \bibinfo
  {author} {\bibfnamefont {K.-i.}\ \bibnamefont {Tainaka}},\ }\href@noop {}
  {\bibfield  {journal} {\bibinfo  {journal} {Journal of the Physical Society
  of Japan}\ }\textbf {\bibinfo {volume} {87}},\ \bibinfo {pages} {014801}
  (\bibinfo {year} {2018})}\BibitemShut {NoStop}%
\bibitem [{\citenamefont {Ke}\ and\ \citenamefont {Wan}(2020)}]{ke2020effects}%
  \BibitemOpen
  \bibfield  {author} {\bibinfo {author} {\bibfnamefont {P.-J.}\ \bibnamefont
  {Ke}}\ and\ \bibinfo {author} {\bibfnamefont {J.}~\bibnamefont {Wan}},\
  }\href@noop {} {\bibfield  {journal} {\bibinfo  {journal} {Ecological
  Monographs}\ }\textbf {\bibinfo {volume} {90}},\ \bibinfo {pages} {e01391}
  (\bibinfo {year} {2020})}\BibitemShut {NoStop}%
\bibitem [{\citenamefont {Momeni}\ \emph {et~al.}(2017)\citenamefont {Momeni},
  \citenamefont {Xie},\ and\ \citenamefont {Shou}}]{momeni2017lotka}%
  \BibitemOpen
  \bibfield  {author} {\bibinfo {author} {\bibfnamefont {B.}~\bibnamefont
  {Momeni}}, \bibinfo {author} {\bibfnamefont {L.}~\bibnamefont {Xie}}, \ and\
  \bibinfo {author} {\bibfnamefont {W.}~\bibnamefont {Shou}},\ }\href@noop {}
  {\bibfield  {journal} {\bibinfo  {journal} {Elife}\ }\textbf {\bibinfo
  {volume} {6}},\ \bibinfo {pages} {e25051} (\bibinfo {year}
  {2017})}\BibitemShut {NoStop}%
\bibitem [{\citenamefont {Nahum}\ \emph {et~al.}(2011)\citenamefont {Nahum},
  \citenamefont {Harding},\ and\ \citenamefont {Kerr}}]{nahum2011evolution}%
  \BibitemOpen
  \bibfield  {author} {\bibinfo {author} {\bibfnamefont {J.~R.}\ \bibnamefont
  {Nahum}}, \bibinfo {author} {\bibfnamefont {B.~N.}\ \bibnamefont {Harding}},
  \ and\ \bibinfo {author} {\bibfnamefont {B.}~\bibnamefont {Kerr}},\
  }\href@noop {} {\bibfield  {journal} {\bibinfo  {journal} {Proceedings of the
  National Academy of Sciences}\ }\textbf {\bibinfo {volume} {108}},\ \bibinfo
  {pages} {10831} (\bibinfo {year} {2011})}\BibitemShut {NoStop}%
\bibitem [{\citenamefont {Cameron}\ \emph {et~al.}(2009)\citenamefont
  {Cameron}, \citenamefont {White},\ and\ \citenamefont
  {Antonovics}}]{cameron2009parasite}%
  \BibitemOpen
  \bibfield  {author} {\bibinfo {author} {\bibfnamefont {D.~D.}\ \bibnamefont
  {Cameron}}, \bibinfo {author} {\bibfnamefont {A.}~\bibnamefont {White}}, \
  and\ \bibinfo {author} {\bibfnamefont {J.}~\bibnamefont {Antonovics}},\
  }\href@noop {} {\bibfield  {journal} {\bibinfo  {journal} {Journal of
  Ecology}\ }\textbf {\bibinfo {volume} {97}},\ \bibinfo {pages} {1311}
  (\bibinfo {year} {2009})}\BibitemShut {NoStop}%
\bibitem [{\citenamefont {Segura}\ \emph {et~al.}(2013)\citenamefont {Segura},
  \citenamefont {Kruk}, \citenamefont {Calliari}, \citenamefont
  {Garc{\'\i}a-Rodriguez}, \citenamefont {Conde}, \citenamefont {Widdicombe},\
  and\ \citenamefont {Fort}}]{segura2013competition}%
  \BibitemOpen
  \bibfield  {author} {\bibinfo {author} {\bibfnamefont {A.}~\bibnamefont
  {Segura}}, \bibinfo {author} {\bibfnamefont {C.}~\bibnamefont {Kruk}},
  \bibinfo {author} {\bibfnamefont {D.}~\bibnamefont {Calliari}}, \bibinfo
  {author} {\bibfnamefont {F.}~\bibnamefont {Garc{\'\i}a-Rodriguez}}, \bibinfo
  {author} {\bibfnamefont {D.}~\bibnamefont {Conde}}, \bibinfo {author}
  {\bibfnamefont {C.}~\bibnamefont {Widdicombe}}, \ and\ \bibinfo {author}
  {\bibfnamefont {H.}~\bibnamefont {Fort}},\ }\href@noop {} {\bibfield
  {journal} {\bibinfo  {journal} {Scientific reports}\ }\textbf {\bibinfo
  {volume} {3}},\ \bibinfo {pages} {1} (\bibinfo {year} {2013})}\BibitemShut
  {NoStop}%
\bibitem [{\citenamefont {Sinervo}\ and\ \citenamefont
  {Lively}(1996)}]{sinervo1996rock}%
  \BibitemOpen
  \bibfield  {author} {\bibinfo {author} {\bibfnamefont {B.}~\bibnamefont
  {Sinervo}}\ and\ \bibinfo {author} {\bibfnamefont {C.~M.}\ \bibnamefont
  {Lively}},\ }\href@noop {} {\bibfield  {journal} {\bibinfo  {journal}
  {Nature}\ }\textbf {\bibinfo {volume} {380}},\ \bibinfo {pages} {240}
  (\bibinfo {year} {1996})}\BibitemShut {NoStop}%
\bibitem [{\citenamefont {Pfeiffer}\ and\ \citenamefont
  {Morley}(2014)}]{pfeiffer2014evolutionary}%
  \BibitemOpen
  \bibfield  {author} {\bibinfo {author} {\bibfnamefont {T.}~\bibnamefont
  {Pfeiffer}}\ and\ \bibinfo {author} {\bibfnamefont {A.}~\bibnamefont
  {Morley}},\ }\href@noop {} {\bibfield  {journal} {\bibinfo  {journal}
  {Frontiers in molecular biosciences}\ }\textbf {\bibinfo {volume} {1}},\
  \bibinfo {pages} {17} (\bibinfo {year} {2014})}\BibitemShut {NoStop}%
\bibitem [{\citenamefont {Jackson}\ and\ \citenamefont
  {Buss}(1975)}]{jackson1975alleopathy}%
  \BibitemOpen
  \bibfield  {author} {\bibinfo {author} {\bibfnamefont {J.}~\bibnamefont
  {Jackson}}\ and\ \bibinfo {author} {\bibfnamefont {L.}~\bibnamefont {Buss}},\
  }\href@noop {} {\bibfield  {journal} {\bibinfo  {journal} {Proceedings of the
  National Academy of Sciences}\ }\textbf {\bibinfo {volume} {72}},\ \bibinfo
  {pages} {5160} (\bibinfo {year} {1975})}\BibitemShut {NoStop}%
\bibitem [{\citenamefont {Baca{\"e}r}(2011)}]{bacaer2011lotka}%
  \BibitemOpen
  \bibfield  {author} {\bibinfo {author} {\bibfnamefont {N.}~\bibnamefont
  {Baca{\"e}r}},\ }in\ \href@noop {} {\emph {\bibinfo {booktitle} {A short
  history of mathematical population dynamics}}}\ (\bibinfo  {publisher}
  {Springer},\ \bibinfo {year} {2011})\ pp.\ \bibinfo {pages}
  {71--76}\BibitemShut {NoStop}%
\bibitem [{\citenamefont {Lotka}(1920)}]{lotka1920analytical}%
  \BibitemOpen
  \bibfield  {author} {\bibinfo {author} {\bibfnamefont {A.~J.}\ \bibnamefont
  {Lotka}},\ }\href@noop {} {\bibfield  {journal} {\bibinfo  {journal}
  {Proceedings of the National Academy of Sciences}\ }\textbf {\bibinfo
  {volume} {6}},\ \bibinfo {pages} {410} (\bibinfo {year} {1920})}\BibitemShut
  {NoStop}%
\bibitem [{\citenamefont {Volterra}(1926)}]{volterra1926fluctuations}%
  \BibitemOpen
  \bibfield  {author} {\bibinfo {author} {\bibfnamefont {V.}~\bibnamefont
  {Volterra}},\ }\href@noop {} {\enquote {\bibinfo {title} {Fluctuations in the
  abundance of a species considered mathematically 1},}\ } (\bibinfo {year}
  {1926})\BibitemShut {NoStop}%
\bibitem [{\citenamefont {Chi}\ \emph {et~al.}(1998)\citenamefont {Chi},
  \citenamefont {Wu},\ and\ \citenamefont {Hsu}}]{chi1998asymmetric}%
  \BibitemOpen
  \bibfield  {author} {\bibinfo {author} {\bibfnamefont {C.-W.}\ \bibnamefont
  {Chi}}, \bibinfo {author} {\bibfnamefont {L.-I.}\ \bibnamefont {Wu}}, \ and\
  \bibinfo {author} {\bibfnamefont {S.-B.}\ \bibnamefont {Hsu}},\ }\href@noop
  {} {\bibfield  {journal} {\bibinfo  {journal} {SIAM Journal on Applied
  Mathematics}\ }\textbf {\bibinfo {volume} {58}},\ \bibinfo {pages} {211}
  (\bibinfo {year} {1998})}\BibitemShut {NoStop}%
\bibitem [{\citenamefont {Frey}(2010)}]{frey2010evolutionary}%
  \BibitemOpen
  \bibfield  {author} {\bibinfo {author} {\bibfnamefont {E.}~\bibnamefont
  {Frey}},\ }\href@noop {} {\bibfield  {journal} {\bibinfo  {journal} {Physica
  A: Statistical Mechanics and its Applications}\ }\textbf {\bibinfo {volume}
  {389}},\ \bibinfo {pages} {4265} (\bibinfo {year} {2010})}\BibitemShut
  {NoStop}%
\bibitem [{\citenamefont {M{\"u}ller}\ and\ \citenamefont
  {Gallas}(2010)}]{muller2010community}%
  \BibitemOpen
  \bibfield  {author} {\bibinfo {author} {\bibfnamefont {A.~P.~O.}\
  \bibnamefont {M{\"u}ller}}\ and\ \bibinfo {author} {\bibfnamefont {J.~A.}\
  \bibnamefont {Gallas}},\ }\href@noop {} {\bibfield  {journal} {\bibinfo
  {journal} {Physical Review E}\ }\textbf {\bibinfo {volume} {82}},\ \bibinfo
  {pages} {052901} (\bibinfo {year} {2010})}\BibitemShut {NoStop}%
\bibitem [{\citenamefont {Szolnoki}\ and\ \citenamefont
  {Perc}(2015)}]{szolnoki2015vortices}%
  \BibitemOpen
  \bibfield  {author} {\bibinfo {author} {\bibfnamefont {A.}~\bibnamefont
  {Szolnoki}}\ and\ \bibinfo {author} {\bibfnamefont {M.}~\bibnamefont
  {Perc}},\ }\href@noop {} {\bibfield  {journal} {\bibinfo  {journal} {New
  Journal of Physics}\ }\textbf {\bibinfo {volume} {17}},\ \bibinfo {pages}
  {113033} (\bibinfo {year} {2015})}\BibitemShut {NoStop}%
\bibitem [{\citenamefont {Kelsic}\ \emph {et~al.}(2015)\citenamefont {Kelsic},
  \citenamefont {Zhao}, \citenamefont {Vetsigian},\ and\ \citenamefont
  {Kishony}}]{kelsic2015counteraction}%
  \BibitemOpen
  \bibfield  {author} {\bibinfo {author} {\bibfnamefont {E.~D.}\ \bibnamefont
  {Kelsic}}, \bibinfo {author} {\bibfnamefont {J.}~\bibnamefont {Zhao}},
  \bibinfo {author} {\bibfnamefont {K.}~\bibnamefont {Vetsigian}}, \ and\
  \bibinfo {author} {\bibfnamefont {R.}~\bibnamefont {Kishony}},\ }\href@noop
  {} {\bibfield  {journal} {\bibinfo  {journal} {Nature}\ }\textbf {\bibinfo
  {volume} {521}},\ \bibinfo {pages} {516} (\bibinfo {year}
  {2015})}\BibitemShut {NoStop}%
\bibitem [{\citenamefont {Guo}\ \emph {et~al.}(2020)\citenamefont {Guo},
  \citenamefont {Song}, \citenamefont {Ge{\v{c}}ek}, \citenamefont {Li},
  \citenamefont {Jusup}, \citenamefont {Perc}, \citenamefont {Moreno},
  \citenamefont {Boccaletti},\ and\ \citenamefont {Wang}}]{guo2020novel}%
  \BibitemOpen
  \bibfield  {author} {\bibinfo {author} {\bibfnamefont {H.}~\bibnamefont
  {Guo}}, \bibinfo {author} {\bibfnamefont {Z.}~\bibnamefont {Song}}, \bibinfo
  {author} {\bibfnamefont {S.}~\bibnamefont {Ge{\v{c}}ek}}, \bibinfo {author}
  {\bibfnamefont {X.}~\bibnamefont {Li}}, \bibinfo {author} {\bibfnamefont
  {M.}~\bibnamefont {Jusup}}, \bibinfo {author} {\bibfnamefont
  {M.}~\bibnamefont {Perc}}, \bibinfo {author} {\bibfnamefont {Y.}~\bibnamefont
  {Moreno}}, \bibinfo {author} {\bibfnamefont {S.}~\bibnamefont {Boccaletti}},
  \ and\ \bibinfo {author} {\bibfnamefont {Z.}~\bibnamefont {Wang}},\
  }\href@noop {} {\bibfield  {journal} {\bibinfo  {journal} {Journal of the
  Royal Society Interface}\ }\textbf {\bibinfo {volume} {17}},\ \bibinfo
  {pages} {20190789} (\bibinfo {year} {2020})}\BibitemShut {NoStop}%
\bibitem [{\citenamefont {Reichenbach}\ \emph {et~al.}(2006)\citenamefont
  {Reichenbach}, \citenamefont {Mobilia},\ and\ \citenamefont
  {Frey}}]{reichenbach2006coexistence}%
  \BibitemOpen
  \bibfield  {author} {\bibinfo {author} {\bibfnamefont {T.}~\bibnamefont
  {Reichenbach}}, \bibinfo {author} {\bibfnamefont {M.}~\bibnamefont
  {Mobilia}}, \ and\ \bibinfo {author} {\bibfnamefont {E.}~\bibnamefont
  {Frey}},\ }\href@noop {} {\bibfield  {journal} {\bibinfo  {journal} {Physical
  Review E}\ }\textbf {\bibinfo {volume} {74}},\ \bibinfo {pages} {051907}
  (\bibinfo {year} {2006})}\BibitemShut {NoStop}%
\bibitem [{\citenamefont {Serrao}\ \emph {et~al.}(2021)\citenamefont {Serrao},
  \citenamefont {Ritmeester},\ and\ \citenamefont
  {Meyer-Ortmanns}}]{serrao2021rare}%
  \BibitemOpen
  \bibfield  {author} {\bibinfo {author} {\bibfnamefont {S.~R.}\ \bibnamefont
  {Serrao}}, \bibinfo {author} {\bibfnamefont {T.}~\bibnamefont {Ritmeester}},
  \ and\ \bibinfo {author} {\bibfnamefont {H.}~\bibnamefont {Meyer-Ortmanns}},\
  }\href@noop {} {\bibfield  {journal} {\bibinfo  {journal} {Journal of Physics
  A: Mathematical and Theoretical}\ }\textbf {\bibinfo {volume} {54}},\
  \bibinfo {pages} {235001} (\bibinfo {year} {2021})}\BibitemShut {NoStop}%
\bibitem [{\citenamefont {West}\ and\ \citenamefont
  {Mobilia}(2020)}]{west2020fixation}%
  \BibitemOpen
  \bibfield  {author} {\bibinfo {author} {\bibfnamefont {R.}~\bibnamefont
  {West}}\ and\ \bibinfo {author} {\bibfnamefont {M.}~\bibnamefont {Mobilia}},\
  }\href@noop {} {\bibfield  {journal} {\bibinfo  {journal} {Journal of
  theoretical biology}\ }\textbf {\bibinfo {volume} {491}},\ \bibinfo {pages}
  {110135} (\bibinfo {year} {2020})}\BibitemShut {NoStop}%
\bibitem [{\citenamefont {Berr}\ \emph {et~al.}(2009)\citenamefont {Berr},
  \citenamefont {Reichenbach}, \citenamefont {Schottenloher},\ and\
  \citenamefont {Frey}}]{berr2009zero}%
  \BibitemOpen
  \bibfield  {author} {\bibinfo {author} {\bibfnamefont {M.}~\bibnamefont
  {Berr}}, \bibinfo {author} {\bibfnamefont {T.}~\bibnamefont {Reichenbach}},
  \bibinfo {author} {\bibfnamefont {M.}~\bibnamefont {Schottenloher}}, \ and\
  \bibinfo {author} {\bibfnamefont {E.}~\bibnamefont {Frey}},\ }\href@noop {}
  {\bibfield  {journal} {\bibinfo  {journal} {Physical review letters}\
  }\textbf {\bibinfo {volume} {102}},\ \bibinfo {pages} {048102} (\bibinfo
  {year} {2009})}\BibitemShut {NoStop}%
\bibitem [{\citenamefont {Bhattacharyya}\ \emph {et~al.}(2020)\citenamefont
  {Bhattacharyya}, \citenamefont {Sinha}, \citenamefont {De},\ and\
  \citenamefont {Hens}}]{bhattacharyya2020mortality}%
  \BibitemOpen
  \bibfield  {author} {\bibinfo {author} {\bibfnamefont {S.}~\bibnamefont
  {Bhattacharyya}}, \bibinfo {author} {\bibfnamefont {P.}~\bibnamefont
  {Sinha}}, \bibinfo {author} {\bibfnamefont {R.}~\bibnamefont {De}}, \ and\
  \bibinfo {author} {\bibfnamefont {C.}~\bibnamefont {Hens}},\ }\href@noop {}
  {\bibfield  {journal} {\bibinfo  {journal} {Physical Review E}\ }\textbf
  {\bibinfo {volume} {102}},\ \bibinfo {pages} {012220} (\bibinfo {year}
  {2020})}\BibitemShut {NoStop}%
\bibitem [{\citenamefont {Reichenbach}\ \emph {et~al.}(2008)\citenamefont
  {Reichenbach}, \citenamefont {Mobilia},\ and\ \citenamefont
  {Frey}}]{reichenbach2008self}%
  \BibitemOpen
  \bibfield  {author} {\bibinfo {author} {\bibfnamefont {T.}~\bibnamefont
  {Reichenbach}}, \bibinfo {author} {\bibfnamefont {M.}~\bibnamefont
  {Mobilia}}, \ and\ \bibinfo {author} {\bibfnamefont {E.}~\bibnamefont
  {Frey}},\ }\href@noop {} {\bibfield  {journal} {\bibinfo  {journal} {Journal
  of Theoretical Biology}\ }\textbf {\bibinfo {volume} {254}},\ \bibinfo
  {pages} {368} (\bibinfo {year} {2008})}\BibitemShut {NoStop}%
\bibitem [{\citenamefont {Kirkup}\ and\ \citenamefont
  {Riley}(2004)}]{kirkup2004antibiotic}%
  \BibitemOpen
  \bibfield  {author} {\bibinfo {author} {\bibfnamefont {B.~C.}\ \bibnamefont
  {Kirkup}}\ and\ \bibinfo {author} {\bibfnamefont {M.~A.}\ \bibnamefont
  {Riley}},\ }\href@noop {} {\bibfield  {journal} {\bibinfo  {journal}
  {Nature}\ }\textbf {\bibinfo {volume} {428}},\ \bibinfo {pages} {412}
  (\bibinfo {year} {2004})}\BibitemShut {NoStop}%
\bibitem [{\citenamefont {Avelino}\ \emph {et~al.}(2012)\citenamefont
  {Avelino}, \citenamefont {Bazeia}, \citenamefont {Losano}, \citenamefont
  {Menezes},\ and\ \citenamefont {Oliveira}}]{avelino2012junctions}%
  \BibitemOpen
  \bibfield  {author} {\bibinfo {author} {\bibfnamefont {P.}~\bibnamefont
  {Avelino}}, \bibinfo {author} {\bibfnamefont {D.}~\bibnamefont {Bazeia}},
  \bibinfo {author} {\bibfnamefont {L.}~\bibnamefont {Losano}}, \bibinfo
  {author} {\bibfnamefont {J.}~\bibnamefont {Menezes}}, \ and\ \bibinfo
  {author} {\bibfnamefont {B.}~\bibnamefont {Oliveira}},\ }\href@noop {}
  {\bibfield  {journal} {\bibinfo  {journal} {Physical Review E}\ }\textbf
  {\bibinfo {volume} {86}},\ \bibinfo {pages} {036112} (\bibinfo {year}
  {2012})}\BibitemShut {NoStop}%
\bibitem [{\citenamefont {Avelino}\ \emph {et~al.}(2018)\citenamefont
  {Avelino}, \citenamefont {Bazeia}, \citenamefont {Losano}, \citenamefont
  {Menezes}, \citenamefont {de~Oliveira},\ and\ \citenamefont
  {Santos}}]{avelino2018directional}%
  \BibitemOpen
  \bibfield  {author} {\bibinfo {author} {\bibfnamefont {P.}~\bibnamefont
  {Avelino}}, \bibinfo {author} {\bibfnamefont {D.}~\bibnamefont {Bazeia}},
  \bibinfo {author} {\bibfnamefont {L.}~\bibnamefont {Losano}}, \bibinfo
  {author} {\bibfnamefont {J.}~\bibnamefont {Menezes}}, \bibinfo {author}
  {\bibfnamefont {B.}~\bibnamefont {de~Oliveira}}, \ and\ \bibinfo {author}
  {\bibfnamefont {M.}~\bibnamefont {Santos}},\ }\href@noop {} {\bibfield
  {journal} {\bibinfo  {journal} {Physical Review E}\ }\textbf {\bibinfo
  {volume} {97}},\ \bibinfo {pages} {032415} (\bibinfo {year}
  {2018})}\BibitemShut {NoStop}%
\bibitem [{\citenamefont {Mooney}(1997)}]{mooney1997monte}%
  \BibitemOpen
  \bibfield  {author} {\bibinfo {author} {\bibfnamefont {C.~Z.}\ \bibnamefont
  {Mooney}},\ }\href@noop {} {\emph {\bibinfo {title} {Monte carlo
  simulation}}},\ \bibinfo {number} {116}\ (\bibinfo  {publisher} {Sage},\
  \bibinfo {year} {1997})\BibitemShut {NoStop}%
\bibitem [{\citenamefont {Zio}(2013)}]{zio2013monte}%
  \BibitemOpen
  \bibfield  {author} {\bibinfo {author} {\bibfnamefont {E.}~\bibnamefont
  {Zio}},\ }in\ \href@noop {} {\emph {\bibinfo {booktitle} {The Monte Carlo
  simulation method for system reliability and risk analysis}}}\ (\bibinfo
  {publisher} {Springer},\ \bibinfo {year} {2013})\ pp.\ \bibinfo {pages}
  {19--58}\BibitemShut {NoStop}%
\bibitem [{\citenamefont {Reichenbach}\ \emph
  {et~al.}(2007{\natexlab{b}})\citenamefont {Reichenbach}, \citenamefont
  {Mobilia},\ and\ \citenamefont {Frey}}]{reichenbach2007noise}%
  \BibitemOpen
  \bibfield  {author} {\bibinfo {author} {\bibfnamefont {T.}~\bibnamefont
  {Reichenbach}}, \bibinfo {author} {\bibfnamefont {M.}~\bibnamefont
  {Mobilia}}, \ and\ \bibinfo {author} {\bibfnamefont {E.}~\bibnamefont
  {Frey}},\ }\href@noop {} {\bibfield  {journal} {\bibinfo  {journal} {Physical
  review letters}\ }\textbf {\bibinfo {volume} {99}},\ \bibinfo {pages}
  {238105} (\bibinfo {year} {2007}{\natexlab{b}})}\BibitemShut {NoStop}%
\bibitem [{\citenamefont {Ipsen}\ \emph
  {et~al.}(2000{\natexlab{a}})\citenamefont {Ipsen}, \citenamefont {Hynne},\
  and\ \citenamefont {S{\o}rensen}}]{ipsen2000amplitude}%
  \BibitemOpen
  \bibfield  {author} {\bibinfo {author} {\bibfnamefont {M.}~\bibnamefont
  {Ipsen}}, \bibinfo {author} {\bibfnamefont {F.}~\bibnamefont {Hynne}}, \ and\
  \bibinfo {author} {\bibfnamefont {P.}~\bibnamefont {S{\o}rensen}},\
  }\href@noop {} {\bibfield  {journal} {\bibinfo  {journal} {Physica D:
  Nonlinear Phenomena}\ }\textbf {\bibinfo {volume} {136}},\ \bibinfo {pages}
  {66} (\bibinfo {year} {2000}{\natexlab{a}})}\BibitemShut {NoStop}%
\bibitem [{\citenamefont {Ipsen}\ \emph
  {et~al.}(2000{\natexlab{b}})\citenamefont {Ipsen}, \citenamefont {Kramer},\
  and\ \citenamefont {S{\o}rensen}}]{ipsen2000amplitudee}%
  \BibitemOpen
  \bibfield  {author} {\bibinfo {author} {\bibfnamefont {M.}~\bibnamefont
  {Ipsen}}, \bibinfo {author} {\bibfnamefont {L.}~\bibnamefont {Kramer}}, \
  and\ \bibinfo {author} {\bibfnamefont {P.~G.}\ \bibnamefont {S{\o}rensen}},\
  }\href@noop {} {\bibfield  {journal} {\bibinfo  {journal} {Physics Reports}\
  }\textbf {\bibinfo {volume} {337}},\ \bibinfo {pages} {193} (\bibinfo {year}
  {2000}{\natexlab{b}})}\BibitemShut {NoStop}%
\bibitem [{\citenamefont {Gong}\ and\ \citenamefont
  {Christini}(2003)}]{gong2003antispiral}%
  \BibitemOpen
  \bibfield  {author} {\bibinfo {author} {\bibfnamefont {Y.}~\bibnamefont
  {Gong}}\ and\ \bibinfo {author} {\bibfnamefont {D.~J.}\ \bibnamefont
  {Christini}},\ }\href@noop {} {\bibfield  {journal} {\bibinfo  {journal}
  {Physical review letters}\ }\textbf {\bibinfo {volume} {90}},\ \bibinfo
  {pages} {088302} (\bibinfo {year} {2003})}\BibitemShut {NoStop}%
\bibitem [{\citenamefont {Mondal}\ \emph {et~al.}(2019)\citenamefont {Mondal},
  \citenamefont {Sharma}, \citenamefont {Upadhyay}, \citenamefont
  {Aziz-Alaoui}, \citenamefont {Kundu},\ and\ \citenamefont
  {Hens}}]{mondal2019diffusion}%
  \BibitemOpen
  \bibfield  {author} {\bibinfo {author} {\bibfnamefont {A.}~\bibnamefont
  {Mondal}}, \bibinfo {author} {\bibfnamefont {S.~K.}\ \bibnamefont {Sharma}},
  \bibinfo {author} {\bibfnamefont {R.~K.}\ \bibnamefont {Upadhyay}}, \bibinfo
  {author} {\bibfnamefont {M.}~\bibnamefont {Aziz-Alaoui}}, \bibinfo {author}
  {\bibfnamefont {P.}~\bibnamefont {Kundu}}, \ and\ \bibinfo {author}
  {\bibfnamefont {C.}~\bibnamefont {Hens}},\ }\href@noop {} {\bibfield
  {journal} {\bibinfo  {journal} {Physical Review E}\ }\textbf {\bibinfo
  {volume} {99}},\ \bibinfo {pages} {042307} (\bibinfo {year}
  {2019})}\BibitemShut {NoStop}%
\bibitem [{\citenamefont {Szczesny}\ \emph {et~al.}(2014)\citenamefont
  {Szczesny}, \citenamefont {Mobilia},\ and\ \citenamefont
  {Rucklidge}}]{szczesny2014characterization}%
  \BibitemOpen
  \bibfield  {author} {\bibinfo {author} {\bibfnamefont {B.}~\bibnamefont
  {Szczesny}}, \bibinfo {author} {\bibfnamefont {M.}~\bibnamefont {Mobilia}}, \
  and\ \bibinfo {author} {\bibfnamefont {A.~M.}\ \bibnamefont {Rucklidge}},\
  }\href@noop {} {\bibfield  {journal} {\bibinfo  {journal} {Physical Review
  E}\ }\textbf {\bibinfo {volume} {90}},\ \bibinfo {pages} {032704} (\bibinfo
  {year} {2014})}\BibitemShut {NoStop}%
\bibitem [{\citenamefont {Mondal}\ \emph {et~al.}(2021)\citenamefont {Mondal},
  \citenamefont {Mondal}, \citenamefont {Kumar~Sharma}, \citenamefont
  {Kumar~Upadhyay},\ and\ \citenamefont
  {Antonopoulos}}]{mondal2021spatiotemporal}%
  \BibitemOpen
  \bibfield  {author} {\bibinfo {author} {\bibfnamefont {A.}~\bibnamefont
  {Mondal}}, \bibinfo {author} {\bibfnamefont {A.}~\bibnamefont {Mondal}},
  \bibinfo {author} {\bibfnamefont {S.}~\bibnamefont {Kumar~Sharma}}, \bibinfo
  {author} {\bibfnamefont {R.}~\bibnamefont {Kumar~Upadhyay}}, \ and\ \bibinfo
  {author} {\bibfnamefont {C.~G.}\ \bibnamefont {Antonopoulos}},\ }\href@noop
  {} {\bibfield  {journal} {\bibinfo  {journal} {Chaos: An Interdisciplinary
  Journal of Nonlinear Science}\ }\textbf {\bibinfo {volume} {31}},\ \bibinfo
  {pages} {103122} (\bibinfo {year} {2021})}\BibitemShut {NoStop}%
\bibitem [{\citenamefont {He}\ \emph {et~al.}(2011)\citenamefont {He},
  \citenamefont {Mobilia},\ and\ \citenamefont
  {T{\"a}uber}}]{he2011coexistence}%
  \BibitemOpen
  \bibfield  {author} {\bibinfo {author} {\bibfnamefont {Q.}~\bibnamefont
  {He}}, \bibinfo {author} {\bibfnamefont {M.}~\bibnamefont {Mobilia}}, \ and\
  \bibinfo {author} {\bibfnamefont {U.~C.}\ \bibnamefont {T{\"a}uber}},\
  }\href@noop {} {\bibfield  {journal} {\bibinfo  {journal} {The European
  Physical Journal B}\ }\textbf {\bibinfo {volume} {82}},\ \bibinfo {pages}
  {97} (\bibinfo {year} {2011})}\BibitemShut {NoStop}%
\bibitem [{\citenamefont {Bazeia}\ \emph {et~al.}(2021)\citenamefont {Bazeia},
  \citenamefont {Bongestab}, \citenamefont {de~Oliveira},\ and\ \citenamefont
  {Szolnoki}}]{bazeia2021effects}%
  \BibitemOpen
  \bibfield  {author} {\bibinfo {author} {\bibfnamefont {D.}~\bibnamefont
  {Bazeia}}, \bibinfo {author} {\bibfnamefont {M.}~\bibnamefont {Bongestab}},
  \bibinfo {author} {\bibfnamefont {B.}~\bibnamefont {de~Oliveira}}, \ and\
  \bibinfo {author} {\bibfnamefont {A.}~\bibnamefont {Szolnoki}},\ }\href@noop
  {} {\bibfield  {journal} {\bibinfo  {journal} {Chaos, Solitons \& Fractals}\
  }\textbf {\bibinfo {volume} {151}},\ \bibinfo {pages} {111255} (\bibinfo
  {year} {2021})}\BibitemShut {NoStop}%
\bibitem [{\citenamefont {de~Oliveira}\ \emph {et~al.}(2021)\citenamefont
  {de~Oliveira}, \citenamefont {de~Moraes}, \citenamefont {Bazeia},\ and\
  \citenamefont {Szolnoki}}]{de2021mobility}%
  \BibitemOpen
  \bibfield  {author} {\bibinfo {author} {\bibfnamefont {B.}~\bibnamefont
  {de~Oliveira}}, \bibinfo {author} {\bibfnamefont {M.}~\bibnamefont
  {de~Moraes}}, \bibinfo {author} {\bibfnamefont {D.}~\bibnamefont {Bazeia}}, \
  and\ \bibinfo {author} {\bibfnamefont {A.}~\bibnamefont {Szolnoki}},\
  }\href@noop {} {\bibfield  {journal} {\bibinfo  {journal} {Physica A:
  Statistical Mechanics and its Applications}\ }\textbf {\bibinfo {volume}
  {572}},\ \bibinfo {pages} {125854} (\bibinfo {year} {2021})}\BibitemShut
  {NoStop}%
\bibitem [{\citenamefont {Drescher}\ \emph {et~al.}(2014)\citenamefont
  {Drescher}, \citenamefont {Nadell}, \citenamefont {Stone}, \citenamefont
  {Wingreen},\ and\ \citenamefont {Bassler}}]{drescher2014solutions}%
  \BibitemOpen
  \bibfield  {author} {\bibinfo {author} {\bibfnamefont {K.}~\bibnamefont
  {Drescher}}, \bibinfo {author} {\bibfnamefont {C.~D.}\ \bibnamefont
  {Nadell}}, \bibinfo {author} {\bibfnamefont {H.~A.}\ \bibnamefont {Stone}},
  \bibinfo {author} {\bibfnamefont {N.~S.}\ \bibnamefont {Wingreen}}, \ and\
  \bibinfo {author} {\bibfnamefont {B.~L.}\ \bibnamefont {Bassler}},\
  }\href@noop {} {\bibfield  {journal} {\bibinfo  {journal} {Current Biology}\
  }\textbf {\bibinfo {volume} {24}},\ \bibinfo {pages} {50} (\bibinfo {year}
  {2014})}\BibitemShut {NoStop}%
\end{thebibliography}%
	\bibliographystyle{apsrev4-1}

\end{document}